\documentclass[letterpaper,12pt]{article}   
\usepackage{osajnl2} 
\usepackage{graphics}
\usepackage{epsfig}
\usepackage{amsmath}

\begin{document}
\title{Intracavity trace molecular detection with a broadband mid-IR frequency comb source}


\author{Magnus~W.~Haakestad,$^{1,2}$ Tobias~P.~Lamour,$^{1,3}$ Nick~Leindecker,$^1$ Alireza~Marandi,$^1$ and Konstantin~L.~Vodopyanov$^{1,4,*}$}
\address{$^1$E. L. Ginzton Laboratory, Stanford University, Stanford, CA 94305, USA}
\address{$^2$Norwegian Defence Research Establishment (FFI), P O Box 25, \\ NO-2027 Kjeller, Norway}
\address{$^3$Scottish Universities Physics Alliance (SUPA), \\Institute of Photonics and Quantum Sciences, School of Engineering and Physical Sciences, \\Heriot-Watt University, Riccarton, Edinburgh EH14 4AS, UK}
\address{$^4$Univ. Central Florida, CREOL, College of Optics \& Photonics, Orlando, FL 32816, USA}
\address{$^*$Corresponding author: vodopyan@stanford.edu}

\begin{abstract}
Ultrasensitive detection of methane, isotopic carbon dioxide, carbon monoxide,
formaldehyde, acetylene and ethylene is performed in the spectral
range 2.5 -- 5 $\mu$m using
intracavity spectroscopy in broadband optical parametric oscillators (OPOs). The OPOs
were operated near degeneracy and synchronously pumped either by a mode-locked erbium (1560
nm) or thulium (2050 nm) fiber laser. A large instantaneous bandwidth of up to 800 cm$^{-1}$ 
allows for simultaneous detection of several gases.
We observe an effective path length enhancement
due to coherent interaction inside the OPO cavity and achieve part-per-billion sensitivity
levels. The measured spectral shapes are in good agreement with a model that takes into
account group delay dispersion across the broad OPO frequency band. 
\end{abstract}


\maketitle 

\section{Introduction}
Many important gases strongly absorb in the mid-IR ($>\negthickspace 2.5\;\mu$m) 'signature' region due to their
fundamental rotational-vibrational transitions in this wavelength range. Optical spectroscopy in
the mid-IR region has potential for such applications as trace gas detection \cite{esler:2000}, remote chemical
sensing \cite{Schliesser:comb}, and human breath analysis \cite{thorpe:2008,ars:2011,risby:breath}. For example, human breath
is known to contain more than 500 different 'biomarker' volatile organic compounds and quantification of
these gases may have clinical applications.

Simultaneous detection of several gases requires a suitable broadband or a widely tunable CW source. Compared to narrow bandwidth sources, 
a broadband source coupled with Fourier transform methods offers advantages for spectroscopy including massive parallelism of data collection and
elimination of the need for wavelength tuning.  Optical frequency combs are particularly attractive broadband sources for spectroscopy 
\cite{didd:comb}, owing to their extraordinary coherence over broad bandwidth.  This property has led to applications including trace gas 
detection, molecular fingerprinting and dual comb spectroscopy \cite{didd:finger,sor:finger,man:ftir,keil:dual}. Even though the
resolution needed for gas spectroscopy may be much lower than the line spacing of frequency
combs, one can utilize the comb structure to improve the sensitivity of trace gas detection. This
can be done by coupling the frequency comb into a high-finesse Fabry-Perot cavity \cite{thorpe:spec, bern:dual, vaer:ftir},
where effective path lengths of several km can be obtained \cite{folty:spec}.

Synchronously pumped (sync-pumped) optical parametric oscillators represent an
attractive way of generating mid-IR frequency combs suitable for
molecular spectroscopy
\cite{till:meth,adl:opo,adl:ftir_opo,folty:spec}. 
Recently, our group implemented a new method for generating broadband mid-IR
combs, based on a doubly resonant, degenerate sync-pumped OPO, which rigorously downconverts
and augments the spectrum of its pump frequency comb \cite{lein:spopo,lein:Tm_spopo}. Exceptionally large
parametric gain bandwidth at degeneracy combined with extensive cross mixing of comb
components, resulted in extremely broad ($>$ one octave) instantaneous mid-IR bandwidth
extending the wavelength range beyond 6 $\mu$m \cite{lein:Tm_spopo}. 
Here we show that such a broadband source, combined with intracavity spectroscopy, becomes a powerful tool for trace molecular detection.

We perform molecular spectroscopy using two such sources. One source is a
periodically-poled lithium niobate (PPLN) based OPO, pumped at 1.56 $\mu$m by a femtosecond Er-doped
fiber laser \cite{lein:spopo}, and the other source is an orientation-patterned GaAs based OPO, pumped
at 2.05 $\mu$m by a Tm-doped fiber laser \cite{lein:Tm_spopo}. Both OPOs operate near degeneracy (central
wavelength is twice that of the pump) -- to obtain broad instantaneous bandwidth. In this work
the OPO cavity itself is used as an enhancement cavity to increase the
effective path length \cite{baev:intracavity,brunner:spectr,boller:spectr}.
Intracavity spectroscopy of methane, carbon monoxide, formaldehyde and several other
gases is performed by injecting gas directly into the OPO enclosure, or by using an intracavity
gas cell with Brewster windows. We observe significant effective path length enhancement due
to the intracavity action. In addition, we find that the measured spectral line shapes may
have dispersive features. Such features have previously been observed with cavity-enhanced
frequency comb spectroscopy \cite{fol:2011,folty:spec}, and in intracavity spectroscopy with sync-pumped OPOs
and mode-locked lasers \cite{vod:spopo_gaas,lein:Tm_spopo,kal:sol}. The measured spectra are compared to a simple model, based
on the intracavity passive loss and round-trip dispersion, and excellent agreement between theory
and measurements is found.

The remainder of the paper is organized as follows. In Sec. \ref{sec:disp}, we introduce a theory
on how intracavity dispersion affects the spectrum and phases of the OPO comb lines. Section \ref{sec:theory}
presents a simple theory of intracavity spectroscopy with broadband OPOs. In Secs. \ref{sec:setup} and \ref{sec:measurements}
we describe the experimental setup and measurement procedures. Sections \ref{sec:er} and \ref{sec:tm} describe
correspondingly results with the Er- and Tm-pumped systems. Finally, in Sec. \ref{sec:discussion}, we discuss
the results and make conclusions in Sec. \ref{sec:conclusions}.
\section{Intracavity dispersion and OPO comb width}\label{sec:disp}
In the case of femtosecond pumping by a mode-locked laser, the pump and the OPO output are
represented by a manifold of comb lines sharing the same spacing, equal to the
pump pulse repetition frequency $f_\text{rep}$. The pump field is represented by a comb of frequencies
$\nu_{p,n}=f_{\text{CEO},p}+nf_\text{rep}$, while the OPO idler and signal follow $\nu_{i,l}=f_{\text{CEO},i}+lf_\text{rep}$ and
$\nu_{s,m}=f_{\text{CEO},s}+mf_\text{rep}$, correspondingly. Here $l,m$, and $n$ are integers, and $f_\text{CEO}$ is the carrier-envelope offset 
frequency ($0\leq f_\text{CEO}<f_\text{rep}$).

From photon energy conservation it follows that for integers $l$ and $m$, there should be an
integer $n$, such that
\begin{equation}
\nu_{i,l} +\nu_{s,m} =\nu_{p,n},
\end{equation}
from which it follows that there are two solutions for $f_\text{CEO}$ of the signal and idler:
\begin{subequations}\label{eq:f_ceo}
\begin{eqnarray}
f_{\text{CEO},s} + f_{\text{CEO},i} &=& f_{\text{CEO},p}\\
f_{\text{CEO},s} + f_{\text{CEO},i} &=& f_{\text{CEO},p} + f_\text{rep}
\end{eqnarray}
\end{subequations}
For a degenerate OPO, where the signal and idler become indistinguishable, $f_{\text{CEO},s} = f_{\text{CEO},i},$ hence
Eq. (\ref{eq:f_ceo}) becomes
\begin{subequations}
\begin{eqnarray}
f_{\text{CEO},s} &=& \frac{f_{\text{CEO},p}}{2},\\
f_{\text{CEO},s} &=& \frac{f_{\text{CEO},p}}{2}+\frac{f_\text{rep}}{2}.\label{eq:al_modes}
\end{eqnarray}
\end{subequations}
In order for a broadband degenerate OPO to oscillate, it is necessary that for each OPO comb frequency, the accumulated phase
delay per roundtrip, $\Delta\phi=kL$, is an integer of $2\pi$ (here $k$ is the average value of the wavevector and $L$ is the
roundtrip cavity length). This can be achieved in vacuum ($k=2\pi\nu/c$) by choosing $L=c/f_\text{rep}$ when $f_{\text{CEO},p}=0$ 
($c$ corresponds to the speed of light). 
\begin{figure}[h!]
\epsfxsize=8.4cm
\centerline{\epsfbox{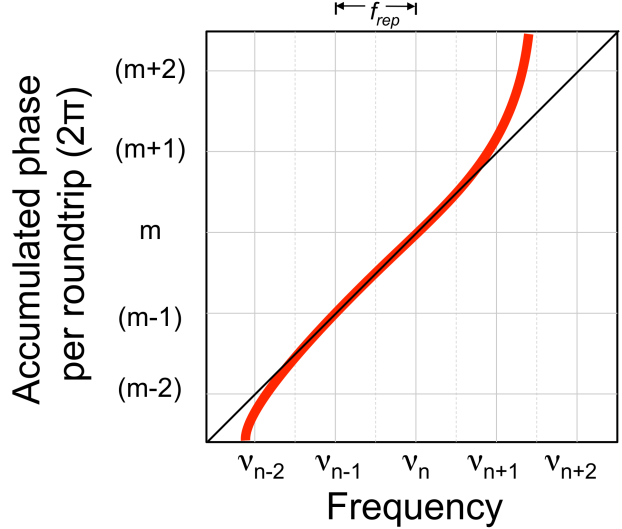}}
\caption{For a broadband OPO to oscillate, it is necessary that for each comb line, $\nu_{s,m}$, the phase delay $kL$ per roundtrip is
an integer multiple of $2\pi$. Due to dispersion in the cavity, the roundtrip phase (thick line) is no longer a straight line. As a
result, peripheral comb lines acquire extra phase, which prevents them from oscillation. Dashed vertical lines correspond to the
alternative set of OPO comb lines shifted in frequency by $f_\text{rep}/2$. The frequency spacing between the comb lines is highly exaggerated 
for clarity.}
\label{fig:acc_phase}
\end{figure}
However due to dispersion in a real cavity, $\Delta\phi$ is no longer
linear with $\nu$ (Fig. \ref{fig:acc_phase}, thick line) and peripheral comb teeth acquire extra phase preventing them from
oscillation.

\begin{figure}[h!]
\epsfxsize=8.4cm
\centerline{\epsfbox{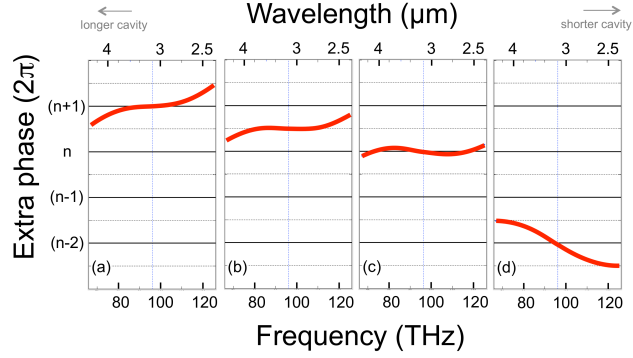}}
\caption{Calculated extra phase per roundtrip near OPO degeneracy (vertical dotted line, 96 THz, 3.12 $\mu$m) at
different cavity roundtrip lengths $L$ near the optimum. The length steps between (a)-(b) and (b)-(c) are 1.56 $\mu$m ($\approx\lambda_\text{pump}$).
Dispersion originating from a 0.5-mm PPLN, 2-mm ZnSe, and a dielectric mirror was taken into account. The cavity length is decreasing from (a) to (d).}
\label{fig:phases}
\end{figure}
Figure \ref{fig:phases} plots the calculated extra phase accumulated per roundtrip near the degeneracy of an
erbium (1560 nm) laser pumped PPLN OPO, at different cavity lengths near the optimum
sync-pumped condition. Here we took into account dispersion originating from (i) 0.5-mm
PPLN, (ii) 2-mm ZnSe dispersion compensator, and (iii) dielectric mirror coating. These components have a
group delay dispersion of -287 fs$^2$, 296 fs$^2$, and 35 fs$^2$, respectively, at 3.12~$\mu$m. Effectively, Fig. \ref{fig:phases} is a residue
between the two curves of Fig. \ref{fig:acc_phase}. When $L$ is decreased (from left to right in Fig. \ref{fig:phases}), the
phase curve rotates clockwise. Broadband OPO operation occurs when the accumulated
phase for all the comb lines is an integer multiple of $2\pi$. Because there is an alternative set of OPO comb lines
Eq. (\ref{eq:al_modes}), the interval between discrete cavity lengths corresponds to $\pi$ instead of $2\pi$, in terms of the
phase shift (this is expressed by dashed horizontal lines), so that the oscillation peaks are spaced
by  $\Delta L\approx\lambda_\text{pump}$ \cite{mar:spopo}. Cases (a)-(c) of Fig. \ref{fig:phases} correspond to degenerate OPO operation, with the broadest spectrum expected for (c), while (d) is a non-degenerate (but still doubly resonant) case,
when the comb lines are clustered around separate signal and idler peaks (at $\sim$70 and 120 THz).

Thus, intracavity dispersion is one of the main factors that restrict the OPO bandwidth
and can also introduce frequency chirp of the OPO pulses.
\section{Model for intracavity molecular spectroscopy}\label{sec:theory}
Now we consider a high-finesse synchronously pumped femtosecond OPO in a steady-state
condition. The round-trip evolution of the electric field $A$ of a resonating OPO comb line at the frequency $\nu_{s,m}$ is
given by
\begin{equation}\label{eq:da}
A = tA + \Delta A.
\end{equation}
Here we assume that $A=A_{s,m}\exp(2\pi i\nu_{s,m}t)$, $t=|t|\exp(-i\Delta\phi)$ is the complex round-trip amplitude transmission
coefficient of the OPO cavity, $\Delta A$ is the nonlinear optical gain due to the presence of the pump
field, and $\Delta\phi$ -- is the phase delay acquired per round-trip. The rationale for using the form Eq. (\ref{eq:da}) is
as follows: For a 3-wave interaction ($\nu_{s,m}+\nu_{i,l}=\nu_{p,n}$) in the slowly varying envelope approximation,
the equation describing the evolution of the OPO field $A(\nu_{s,m})$ in a nonlinear crystal is given by
$\frac{\text{d}}{\text{d}z}A(\nu_{s,m})=i\kappa A(\nu_{p,n})A(\nu_{i,l})^*$, where $\kappa$ is a constant, proportional to the nonlinear 
susceptibility $\chi^{(2)}$, while assuming perfect phase-matching. 
\begin{figure}[h!]
\epsfxsize=8.4cm
\centerline{\epsfbox{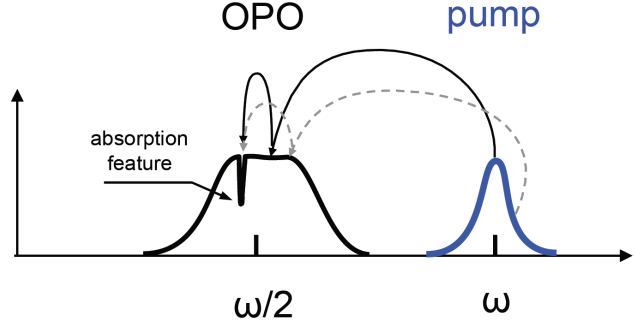}}
\caption{In a broadband OPO, parametric gain at a given frequency is the result of cross-coupling between large
amounts of pump and OPO comb lines. Two such paths are shown.}
\label{fig:cross}
\end{figure}
In the sync-pumped OPO case, the nonlinear interaction results in a cross-coupling between large manifolds
of pump and OPO comb lines (Fig. \ref{fig:cross}), such that for the $m$-th mode
\begin{equation}\label{eq:cross}
\frac{\text{d}}{\text{d}z}A(\nu_{s,m})=i\kappa\sum_{\nu_{p,n}}\sum_{\nu_{i,l}=\nu_{p,n}-\nu_{s,m}}A(\nu_{p,n})A^*(\nu_{i,l}).
\end{equation}
Here $A(\nu_{p,n})$ refers to the pump, $A(\nu_{i,l})$ -- to the 'complementary' OPO wave, and a star denotes its
complex conjugate. Given the extremely broad spectra for both the pump and the OPO, the presence of narrow molecular absorption features in
the spectra will be averaged out in Eq.~(\ref{eq:cross}). Based on this reasoning, we make the
approximation that the right hand side of Eq.~(\ref{eq:cross}) is not significantly affected by weak, narrow molecular absorption features, and integration over the crystal length gives the constant field 'gain' $\Delta A$ in Eq.~(\ref{eq:da}).
\subsection{Intracavity absorption enhancement, real amplitude transmission}
Let us first regard a simple case of high-finesse cavity when the OPO comb lines are in exact
resonance with the cavity, $\exp(-i\Delta\phi)=1$, and the round-trip amplitude transmission $t(\nu)$ is real: 
$t_0=\exp(-\delta_0)$, where 
$\delta_0\ll 1$ is the round-trip amplitude loss, such that $\delta_0=\delta_o^I/2$, where $\delta_0^I$
is the intensity (or power) loss. Also, let us assume that a small \emph{real} (as in the center of a molecular
absorption line) roundtrip absorption $\delta_1$ is introduced inside the cavity, such that $\delta_1\ll\delta_0$,
corresponding to the amplitude transmission coefficient $t_1=\exp(-\delta_1)$. In this case, Eq. (\ref{eq:da}) results in
\begin{equation}\label{eq:amp_tran}
A=\frac{\Delta A}{1-t_0t_1}\approx\frac{\Delta A}{\delta_0+\delta_1}=\frac{\Delta A}{\delta_0(1+\delta_1/\delta_0)}.
\end{equation}
The relative field change $\delta A/A$, due to molecular absorption, is $-\delta_1/(\delta_0+\delta_1)\approx-\delta_1/\delta_0$, instead of 
$-\delta_1$ if one measures transmission change due to molecular gas outside the OPO cavity. Consequently, the
high-finesse OPO cavity 'amplifies' the small absorption by a factor $1/\delta_0$. This can also be expressed
in terms of the photon lifetime inside the cavity. The photon number decays in a cavity as $n\sim n_0\exp(-2\delta_0f_\text{rep}t)$,
and thus $n$ reduces by a factor $1/e$ after $N=1/(2\delta_0)$ cavity roundtrips. 
Hence the absorption enhancement due to the intracavity action, 
which results in a corresponding increase of the effective path length through the gas, is $2N$.
\subsection{Intracavity absorption enhancement, complex amplitude transmission}
\begin{figure}[h!]
\epsfxsize=8.4cm
\centerline{\epsfbox{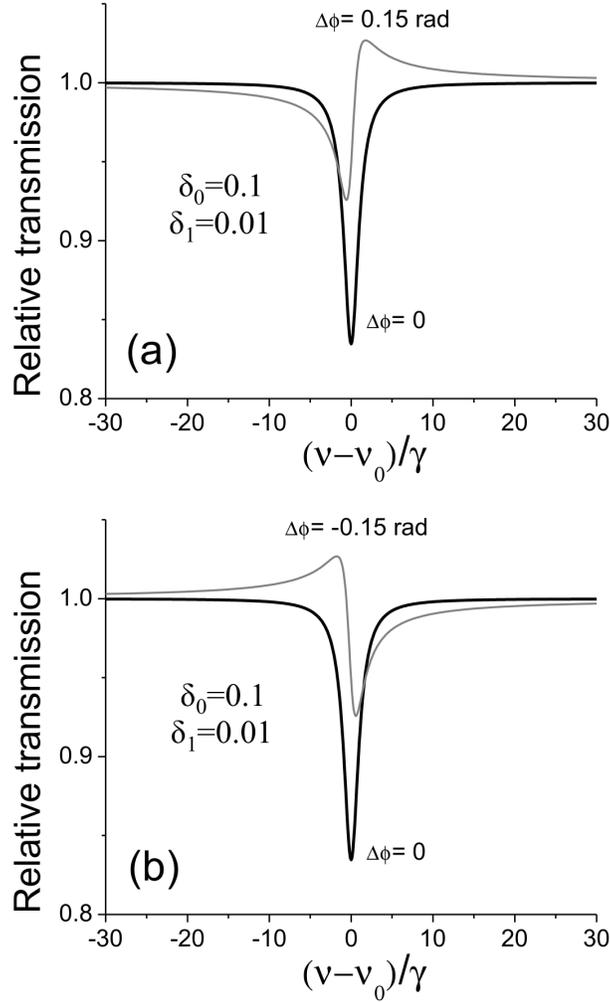}}
\caption{When the roundtrip phase differs from an integer times $2\pi$, dispersive features of both signs appear in the molecular
spectrum. (a) Calculated relative spectral intensity $|A|^2$ near molecular resonance vs. normalized frequency at $\Delta\phi=0$
and $\Delta\phi=0.15$. (b) Same at $\Delta\phi=0$ and $\Delta\phi=-0.15$.}
\label{fig:trans}
\end{figure}

When an OPO comb line acquires extra phase $\Delta\phi$ per round trip, the transmission $t$ is complex: $t_0=\exp(-\delta_0-i\Delta\phi)$. 
Also, molecular absorption (we assume for simplicity a Lorentzian line shape, which is valid at
pressures $>0.1$ atm), can be characterized close to resonance by a complex transmission function \cite{dem:spec}
\begin{equation}\label{eq:t1}
t_1=\exp\left(-\frac{i\delta_1}{(\nu_0-\nu)/\gamma+i}\right).
\end{equation}
Here $\delta_1$ is the \emph{amplitude} loss per roundtrip at the resonant frequency $\nu_0$, and $\gamma$ is the line half width.
For a known gas and known path length $L_\text{mol}$, $\delta_1$ for a given molecular transition can be
found from the HITRAN database through the formula 
$\delta_1=\frac{1}{2}\delta_1^I=\frac{1}{2}\sigma_\text{mol}(\nu_0)n_\text{mol}L_\text{mol}=\frac{1}{2}\frac{S}{\pi\gamma}n_\text{mol}L_\text{mol}$, 
where $S$ is the pressure-independent \emph{absorption intensity} for a given energy transition, that is absorption cross-section per molecule 
integrated over the frequency \cite{hitr:harvard},
$\sigma_\text{mol}(\nu_0)$ is the peak absorption cross-section, and $n_\text{mol}$ is molecular concentration.
Similar to Eq. (\ref{eq:amp_tran}), we can write an expression for the field amplitude:
\begin{equation}\label{eq:A}
A=\frac{\Delta A}{1-t_0t_1}\approx\frac{\Delta A}{1-\exp\left(-\delta_0-i\Delta\phi-\frac{i\delta_1}{(\nu_0-\nu)/\gamma+i}\right)}.
\end{equation}
When $\Delta\phi\neq 0$, derivative-like features appear in the spectrum, as
illustrated by Fig. \ref{fig:trans}. Physically, dispersion in the OPO cavity introduces a
mismatch $\Delta\phi(\nu)$ between the OPO frequency comb
and the cavity resonances. Molecular dispersion imposes an additional phase shift near absorption resonances. 
This causes a change of the spectral line shapes depending on the mismatch. 

Accordingly, our simple theory for intracavity spectroscopy with femtosecond OPOs
predicts dispersive features in the absorption spectrum, similar to those observed with frequency
comb spectroscopy enhanced with an external cavity \cite{fol:2011,gian:1999}.
\section{Experimental setup}\label{sec:setup}
\begin{figure}[h!]
\epsfxsize=8.4cm
\centerline{\epsfbox{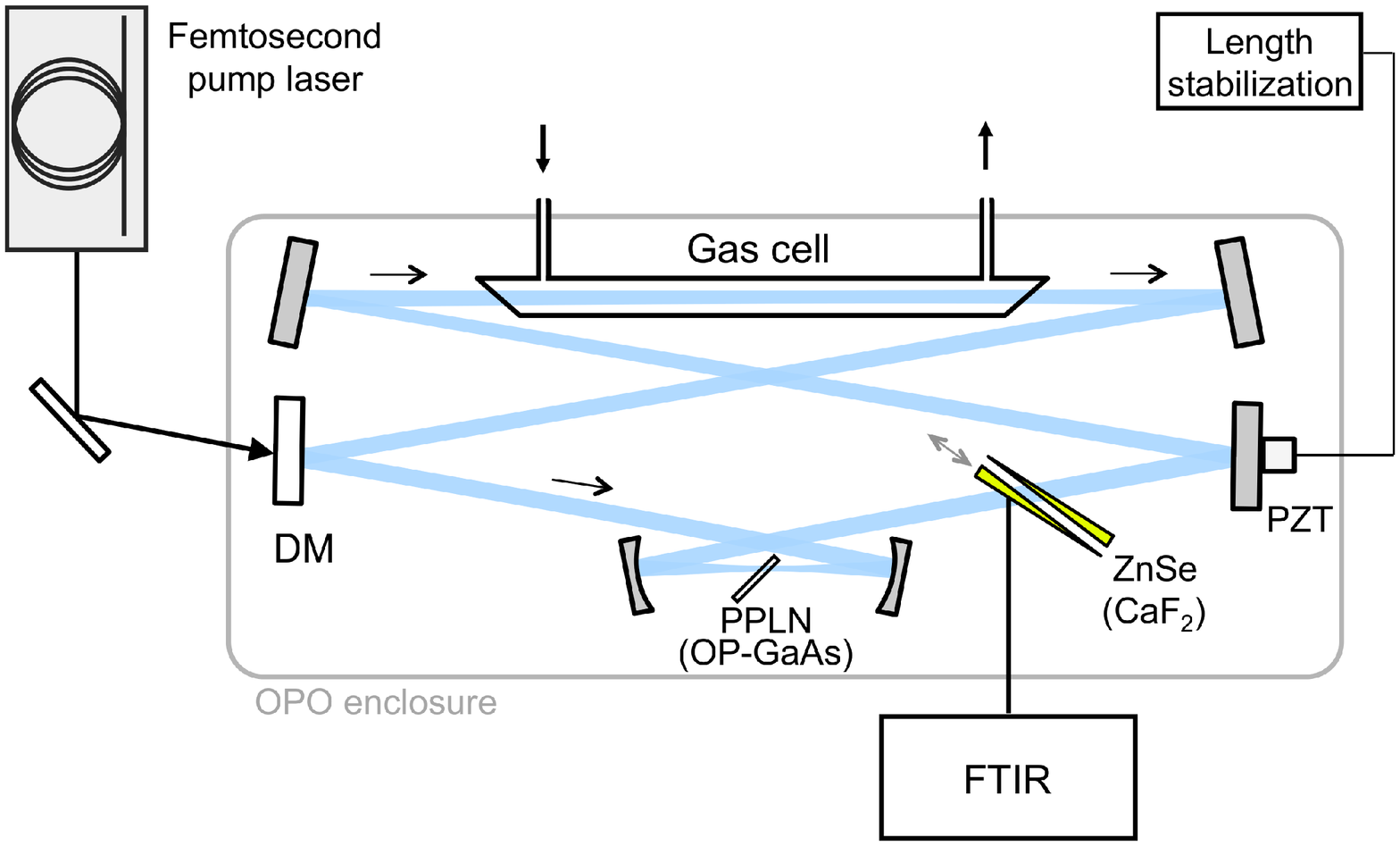}}
\caption{Schematic of the degenerate broadband OPO. The pump beam was introduced through the in-coupling dielectric mirror DM.
The other five mirrors are gold coated. A pair of wedges made of ZnSe (Er:fiber-pumped system) or CaF$_2$ (Tm:fiber-pumped system)
was used for (i) dispersion compensation and (ii) beam out-coupling. The nonlinear crystal was AR-coated (PPLN) or placed at Brewster's angle 
(OP-GaAs).}
\label{fig:setup}
\end{figure}

The first OPO  in our experiment was based on a PPLN crystal and was pumped by an Er:fiber laser (Toptica, 350 mW average power, 1.56 $\mu$m
wavelength, 80 MHz repetition rate, 85 fs pulse duration), producing an output centered at $\sim3.1\;\mu\text{m}$. 
The length of the PPLN crystal was 0.5 mm or 0.8 mm and a ZnSe 1-degree wedge pair was used for
dispersion compensation and for out-coupling, as shown in Fig. \ref{fig:setup}.
The second OPO was based on a 0.5-mm-long orientation-patterned GaAs (OP-GaAs) crystal, and was pumped by a
Tm:fiber laser (IMRA, 600 mW average power, 2.05 $\mu$m wavelength, 75 MHz rep. rate, 93 fs pulse
duration), producing an output centered at 4.1 $\mu$m. The OPOs and their coherence properties are
described in detail in Refs. \cite{lein:spopo,mar:spopo,lein:Tm_spopo}. Both OPOs were placed in Plexiglas enclosures and the
pump lasers were free running. The output power of the OPOs was some tens of mW.

\begin{figure}[h!]
\epsfxsize=8.4cm
\centerline{\epsfbox{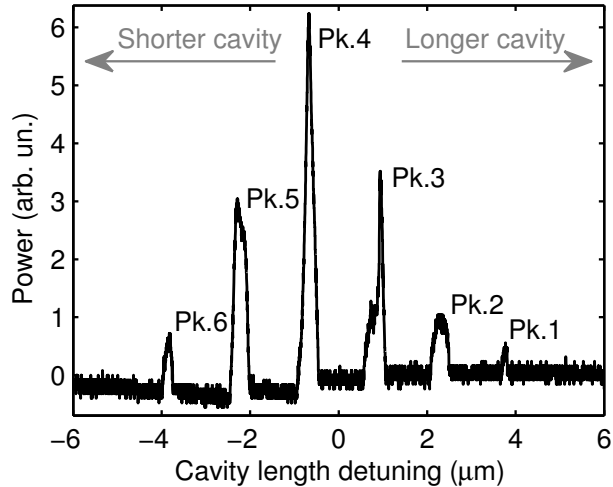}}
\caption{Output power of the Er:fiber pumped OPO, with 0.8 mm PPLN, as a function of cavity length (roundtrip) detuning.}
\label{fig:osc}
\end{figure}
Oscillation occurs at a discrete set of cavity lengths, separated (in effective roundtrip cavity length) by
approximately one pump wavelength, due to the doubly resonant operation \cite{mar:spopo}. Figure \ref{fig:osc} shows measured
output power of the Er:fiber pumped OPO, with 0.8 mm PPLN, as a function of cavity length detuning. The cavity
length was locked to one of these oscillation peaks using a feedback loop including a piezo stage
attached to one of the cavity mirrors. 
\begin{figure}[h!]
\epsfxsize=8.4cm
\centerline{\epsfbox{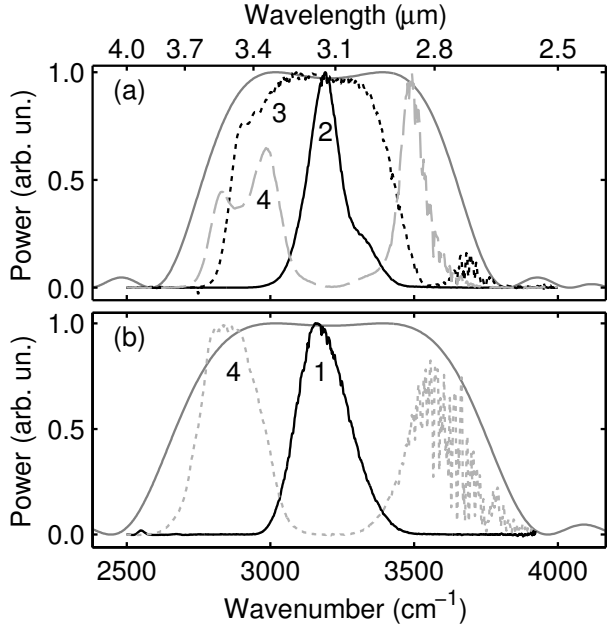}}
\caption{Measured output spectra for the Er:fiber pumped OPO using (a) 0.8 mm PPLN and (b) 0.5 mm PPLN. The
spectra are shown for oscillation peaks No. 2--4 in (a) and peaks No. 1 and 4 in (b). Also shown (solid gray lines) are the theoretical PPLN 
parametric gain spectra for the two crystal lengths.}
\label{fig:spec}
\end{figure}
Figure \ref{fig:spec} shows measured spectra for the Er:fiber OPO for a
selection of oscillation peaks, with N$_2$-purging of the OPO enclosure. The spectra are shown for
the two different PPLN lengths used in the experiments. We observe that the maximum spectral
width is limited by the parametric gain bandwidth of the nonlinear PPLN crystal. The shape of the OPO spectra is strongly
dependent on (i) intracavity dispersion and (ii) on which oscillation peak the OPO is locked to
\cite{lein:Tm_spopo}. 
\begin{figure}[h!]
\epsfxsize=8.4cm
\centerline{\epsfbox{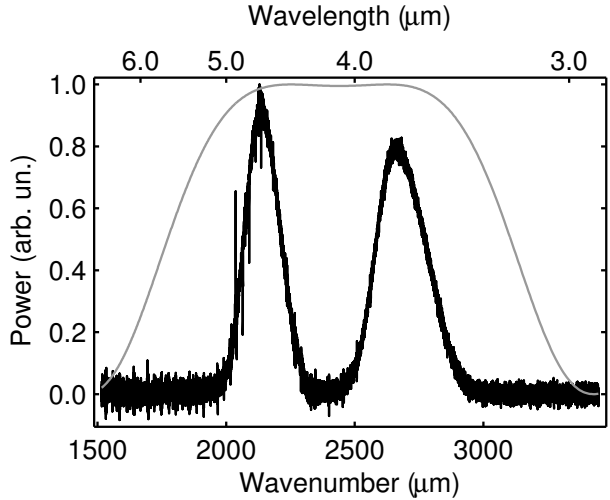}}
\caption{Measured spectrum for one oscillation peak for the Tm:fiber-pumped OPO. Also shown (gray line) is the theoretical gain
spectrum of the 500 $\mu$m OP-GaAs crystal.}
\label{fig:Tm_spec}
\end{figure}
Figure \ref{fig:Tm_spec} shows measured spectrum for one of the oscillation peaks for Tm:fiber pumped
OPO. 
\section{Measurement}\label{sec:measurements}
We have performed spectroscopic measurements of six different molecules. Measurements of
methane, formaldehyde, ethylene and acetylene were performed using the Er:fiber pumped OPO,
while the absorption spectra of carbon monoxide and isotopic carbon dioxide ($^{13}\text{CO}_2$) were
measured with the Tm:fiber pumped OPO. Measurements of the absorption spectra were carried
out by injecting a controlled amount of gas directly into the N$_2$-purged Plexiglas enclosure, or by
using an intra-cavity gas cell with length 48 cm and a volume of $\sim 30$ cm$^3$ (used for formaldehyde and
carbon monoxide). The Brewster windows in the gas cell were made of 1-mm-thick ZnS that was
chosen because of its comparatively low dispersion in the 2--5 $\mu$m range (with zero-dispersion at
3.7 $\mu$m).

The main advantages with injecting gas directly into the OPO enclosure were simplicity
and the ability to exploit the entire cavity round-trip length as a physical path length through the
gas. The disadvantages with this approach were the large volume of the Plexiglas enclosure (estimated to be 103 liter) 
and possible contamination of the optical surfaces of the OPO components. The
advantages using the gas cell were the small volume of the cell, and the ability to easily control
the flow rate and the pressure of the gas. The main disadvantage with the gas cell was the short
cell length compared to the cavity round-trip distance ($\sim 0.5$ m vs. $\sim 4$ m), but this could in principle
be improved by using a different resonator geometry allowing longer intracavity gas cells.

All spectra were measured at 1 atm. pressure and a temperature of 21.5 $^\circ$C. The flow-rate
of gas through the intra-cavity cell was 0.2 l/min. for formaldehyde, and a steady-state gas
concentration was reached after $\sim 5$ min. of flowing. No flowing was used for carbon monoxide.
Unless otherwise noted, the OPO output spectra were measured with a commercial (Nicolet 6700) FTIR spectrometer
with a liquid N$_2$-cooled HgCdTe detector. We used the maximum available resolution of 0.125~cm$^{-1}$ 
of the FTIR instrument for most of the measurements. The total measurement time was in
the range of 30 seconds up to 4 minutes, which includes averaging over 8 to 32 scans. First, the OPO was locked to an
appropriate oscillation peak during measurements using the dither-and-lock technique. Then reference spectra were obtained by filling the
OPO enclosure and gas cell, respectively, with N$_2$ only. Finally, the trace gases were injected and the absorption spectra were determined from the ratio
between these two measurements. The free-running pump lasers resulted in drifts of the OPO
spectra on a time scale of the order of minutes, affecting the baseline in the relative OPO spectra.
We therefore apply a baseline correction (determined by parts of the spectra between the
absorption lines) of up to a few percent of the measured spectral intensity. In addition, an offset in
wavenumber of up to $\sim 1$ cm$^{-1}$ was applied to the measured data, to compensate FTIR instrument errors and to match the positions of the
measured absorption peaks to the HITRAN data.
\section{Results obtained with the Er-pumped system}\label{sec:er}
\subsection{Methane (CH$_4$)}
A controlled amount of 0.2--1\% methane in N$_2$ was injected into the N$_2$-flushed OPO
enclosure. Based on the volume of the enclosure, the concentration of
methane in the OPO cavity was estimated to be 8.5 ppm. 
\begin{figure}[h!]
\centerline{\epsfbox{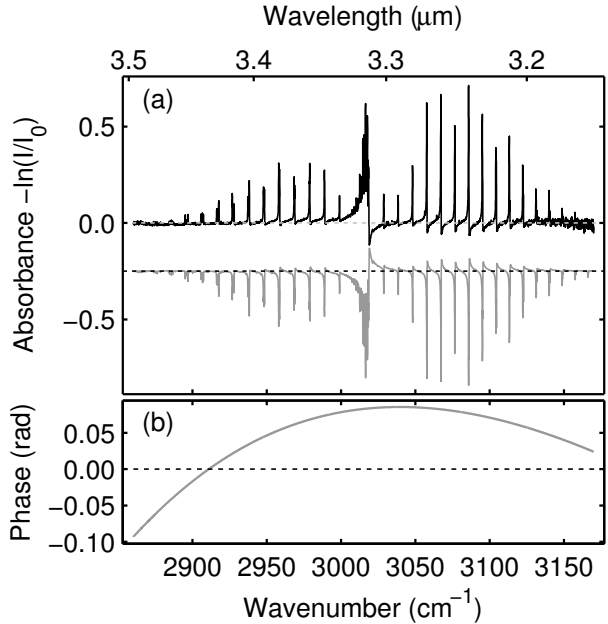}}
\caption{(a) Measured (black) and calculated (gray) absorption spectra for 8.5~ppm methane in nitrogen at 1 atm. pressure, corresponding to peak No. 4 in Fig. \ref{fig:osc}. The calculated spectrum is inverted and offset for clarity. (b) Phase shift $\Delta\phi(\nu)$, which was used for the calculation in (a).}
\label{fig:met}
\end{figure}
Figure \ref{fig:met}(a) shows measured and calculated intracavity spectra, corresponding to absorption of methane
for the OPO using 0.8 mm of PPLN. The measurement time was 57 s. 
The round-trip phase $\Delta\phi$ is determined from the calculated round-trip dispersion of the OPO cavity up to a linear frequency dependence.
We adjust the linear part of $\Delta\phi$ to optimize the fit between the measured and
the calculated spectrum. The calculated absorption of methane was based on the line intensities $S(\nu)$ and line half-widths $\gamma$ from the 
HITRAN database \cite{hitr:harvard}. The resulting complex susceptibility was used as an input to
Eq. (\ref{eq:A}) together with the estimated round-trip phase
$\Delta\phi$, which is shown in Fig.~\ref{fig:met}(b). We estimated the round-trip loss to be 25\%. 
We observe that the simulated spectrum reproduces the measured spectrum well. 
\begin{figure}[h!]
\centerline{\epsfbox{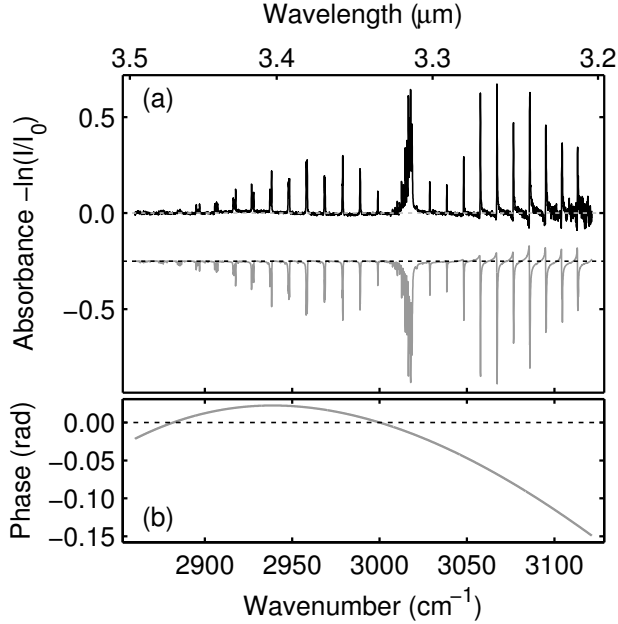}}
\caption{(a) Measured (black) and calculated (gray) absorption spectra for 8.5~ppm
  methane in nitrogen at 1 atm. pressure for an adjacent oscillation
  peak (peak No. 5 in Fig. \ref{fig:osc}), compared to Fig. \ref{fig:met}. The calculated spectrum is inverted and offset for clarity. 
(b) Phase shift $\Delta\phi(\nu)$, which was used for the calculation in (a).}
\label{fig:met_pk5}
\end{figure}
Figure \ref{fig:met_pk5} shows measured and calculated spectra for an
adjacent (in cavity length) OPO oscillation peak (No. 5), compared to Fig. \ref{fig:met}. In the
calculation, we assumed the required cavity length change
between the two oscillation peaks to be 1.71 $\mu$m. Nominally, the change in cavity length between two
adjacent peaks is equal to the pump wavelength \cite{lein:spopo},
which is 1.56 $\mu$m. However, the observed spacing between the oscillation
peaks varies within $\pm 15\%$, which we attribute to intracavity
dispersion.

Based on the measured absorption spectrum and the noise of the detector and the laser source, we estimated a detection
limit of 1.7 ppb for methane with the current experimental parameters. 
Compared to single peak detection, broadband spectroscopy detects multiple absorption peaks simultaneously, which leads to an increase of the 
sensitivity, as pointed out in Ref. \cite{adl:ftir_opo}.
For example, if there are $N$ equally strong absorption peaks, the
multi-line advantage improves the detection limit by a factor $\sqrt{N}$.
For methane, the sensitivity is improved by approximately one order of
magnitude, depending on the spectral resolution \cite{adl:ftir_opo}. In Sec. \ref{sec:discussion}, 
we discuss the procedure of calculating the detection limit in our experiment.

\begin{figure}[h!]
\centerline{\epsfbox{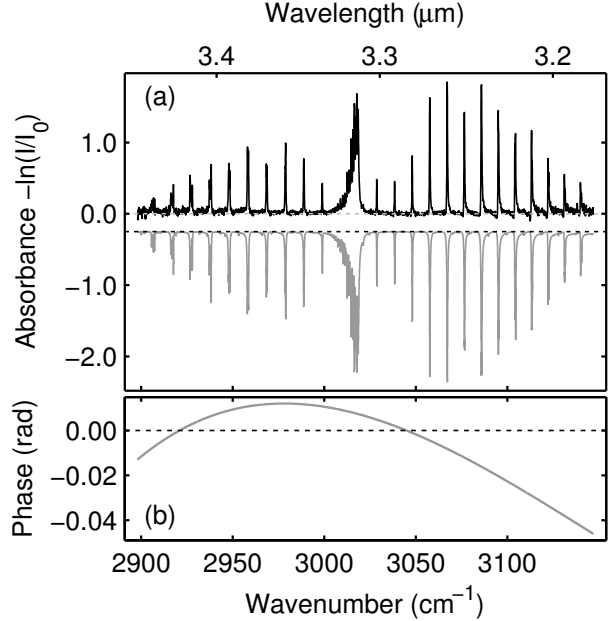}}
\caption{(a) Measured (black) and calculated (gray) absorption spectra for 56 ppm methane in nitrogen at 1 atm. pressure. 
The calculated spectrum is inverted and offset for clarity. (b) Phase shift $\Delta\phi(\nu)$, which was used for the calculation in (a).}
\label{fig:met_56ppm}
\end{figure}
Figure \ref{fig:met_56ppm} shows measured and calculated intracavity spectra for 56 ppm methane, with the Er:fiber-pumped OPO using 0.8 mm PPLN 
and close to zero intracavity dispersion. The estimated round trip loss is about $30\%$, which is due to increased out-coupling, compared to Figs. (\ref{fig:met}) and (\ref{fig:met_pk5}). The spectrum was measured using a DTGS (deuterated triglycine sulfate) detector at room temperature, with 4 min. measurement time. As for Figs. (\ref{fig:met}) and (\ref{fig:met_pk5}), the linear part of the round-trip phase is adjusted to fit the calculated spectrum to the measured spectrum.

\begin{figure}[h!]
\centerline{\epsfbox{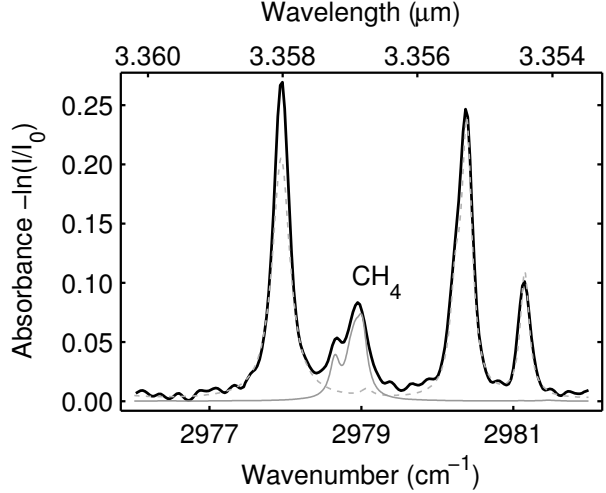}}
\caption{Measured absorption spectrum for ambient air (black line) containing a methane peak. The HITRAN simulation for methane (solid gray line)
and the three simulated (HITRAN) water vapor peaks (dashed gray line) are also shown . The effective path length is taken to be 8 times the round-trip length of the OPO cavity for the calculated spectra.}
\label{fig:air}
\end{figure}
The concentration of methane in standard air is approximately 1.8 ppm,
and many of the methane lines are overlapping with the absorption
lines of water vapor. Figure \ref{fig:air} shows a part of the OPO
spectrum with no N$_2$-flushing of the cavity (relative humidity 33.6\%, temperature 21.5$^\circ$C). The selected part of
the spectrum corresponds to a region where a strong absorption line of
methane is not significantly interfering with absorption lines of water vapor.
We clearly see the absorption feature of methane in air, which is compared in the same figure to a HITRAN simulation for
1.8 ppm of methane and intracavity enhancement factor of 8. A reasonable match between experimental and calculated spectra is consistent
with a round-trip loss of 25\%. Simulated water vapor peaks (HITRAN) for the above atmospheric conditions and enhancement factor are shown
as dashed lines.
\subsection{Formaldehyde (CH$_2$O)}
Measurements of the absorption spectrum of formaldehyde were
carried out using a gas cell with a continuous flow of 100 ppm 
formaldehyde in N$_2$. In these measurements we used a 500-$\mu$m-long
PPLN crystal to obtain a sufficiently wide spectrum, corresponding to
peak 4 in Fig. \ref{fig:spec}(b), to cover the absorption region of
formaldehyde from 3.3 to 3.7 $\mu$m.
\begin{figure}[h!]
\centerline{\epsfbox{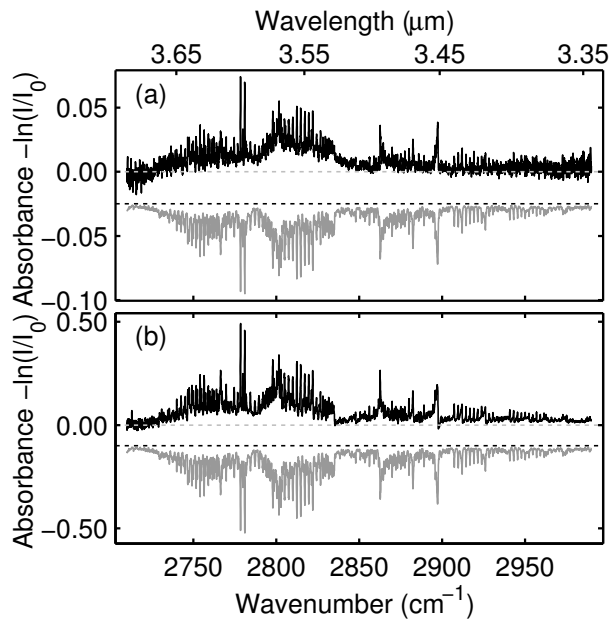}}
\caption{Measured (black) and calculated absorption spectra (gray) for 100 ppm formaldehyde in nitrogen at 1 atm. pressure. The calculated spectra are offset and shown on an inverted scale for clarity. (a) Extracavity spectra, (b) intracavity spectra. The effective path length is taken to be 6 times the length of the gas cell for the calculated spectrum in (b).}
\label{fig:formaldehyde}
\end{figure}
Figure \ref{fig:formaldehyde}(a) and (b) show measured and calculated
spectra of formaldehyde when the gas cell is placed outside (2 min.
measurement time) and inside (30 s measurement time) the OPO cavity, 
respectively. The absorption was enhanced by a factor of about
6 when the gas cell was placed inside the OPO cavity, giving
a detection limit of approximately 0.31 ppm. This enhancement corresponds to an 
intracavity round-trip loss of 33\% for the signal. This loss is
higher than for the measurements with methane, but the increased
round-trip loss may be due to a small misalignment of the gas cell.

For the fitted data we assumed a round-trip phase shift of $\Delta\phi=0$, because this
gave a reasonable overlap with the measured data. 
\subsection{Simultaneous measurement of acetylene (C$_2$H$_2$) and methane (CH$_4$)}\label{sec:Ace_met}
\begin{figure}[h!]
\centerline{\epsfbox{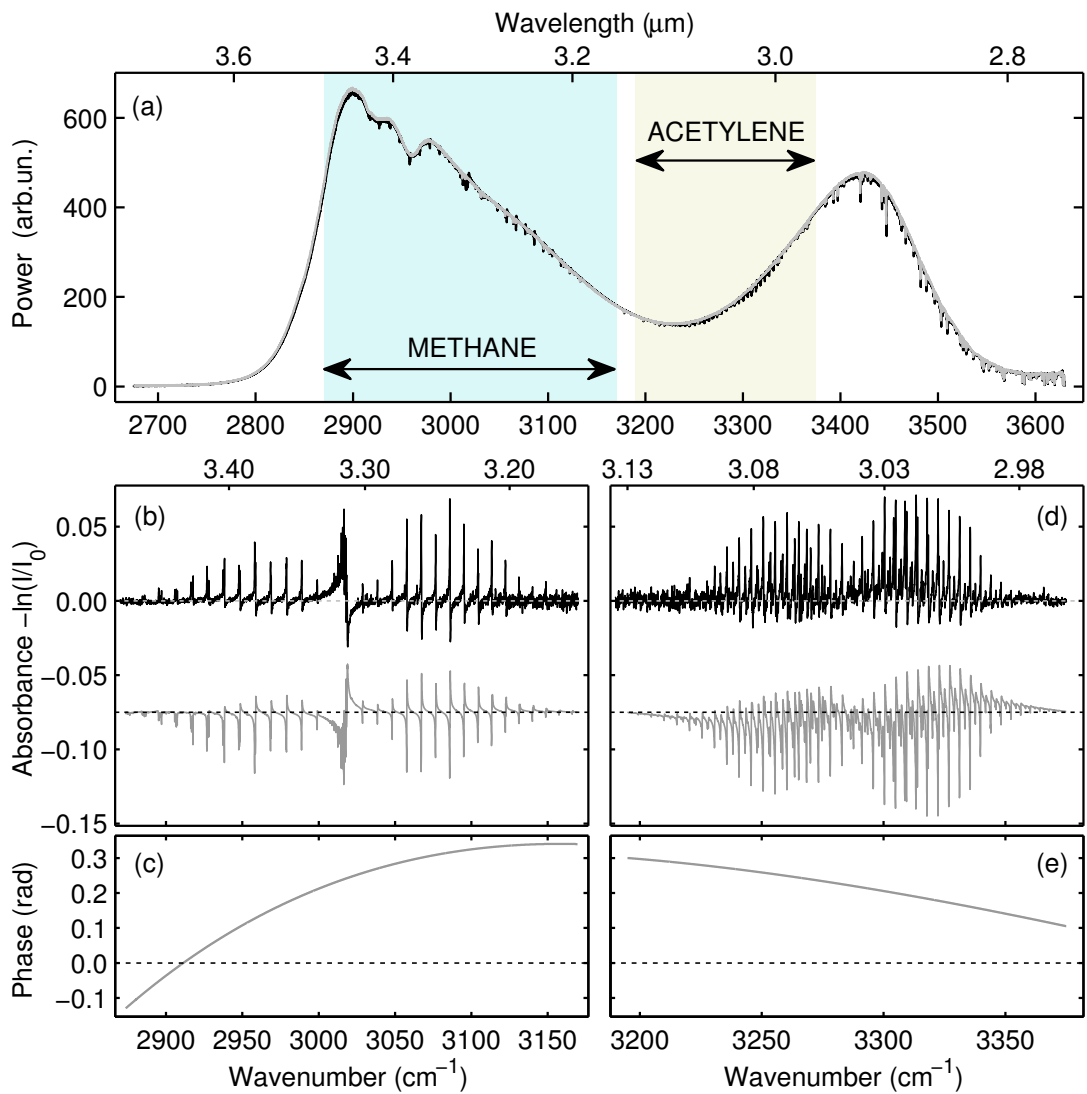}}
\caption{(a) Showing reference intracavity spectrum (gray) and, underneath, the intracavity spectrum (black) with the absorption features
present, while detecting methane and acetylene simultaneously inside the OPO. (b) Experimentally measured methane spectrum (black) at a concentration of 1.4 ppm and the corresponding calculated spectrum (gray). (c) Round-trip phase shift for the calculated methane spectrum. (d) Experimentally measured acetylene spectrum (black) at a concentration of 3.8 ppm and the corresponding calculated spectrum (gray). (e) Round-trip phase shift for the calculated acetylene spectrum. The calculated spectra in (b) and (d) are offset and shown on an inverted scale for clarity.}
\label{fig:Tobi_spec}
\end{figure}

This experiment was performed to measure acetylene on one hand but also to show the capabilities of the intracavity OPO spectroscopy to detect more than one trace gas at once with reasonable high resolution. The use of a syringe allowed a precisely controlled injection of 400 ml of 1000 ppm acetylene in N$_2$, followed by 15.0 ml of 1\% methane in 
N$_2$, into the N$_2$-flushed OPO enclosure, resulting in an estimated 3.8 ppm acetylene concentration and 1.4 ppm methane concentration,
respectively. Figure \ref{fig:Tobi_spec}(a) shows the measured reference spectrum (gray) and, underneath, the absorption intracavity spectrum 
(black), covering the absorption region of methane and acetylene, with the OPO operated with 0.8 mm of PPLN crystal length resonating at its peak
No. 4.
Figure \ref{fig:Tobi_spec}(b) indicates the dispersive absorption spectrum of methane and the corresponding calculated spectrum. Figure \ref{fig:Tobi_spec}(d) contains the detected absorption spectrum for acetylene which shows a similar amount of dispersion. Taking both traces simultaneously did not allow optimizing the spectrum for lowest dispersion across the relevant wavelength ranges while maintaining a good signal to noise ratio for both gases. Setting the FTIR to 17 scans per trace with a resolution of 0.125 cm$^{-1}$ resulted in an acquisition time of 60 s for the reference as well as the absorption measurements. The simulated spectra were obtained using the phase shifts in Fig. \ref{fig:Tobi_spec}(c) and (e) for methane and acetylene, respectively, while assuming a cavity round-trip loss of 25\%. We estimate a detection limit of 0.11 ppm for acetylene with the 
current experimental conditions.
\subsection{Ethylene (C$_2$H$_4$)}
\begin{figure}[h!]
\centerline{\epsfbox{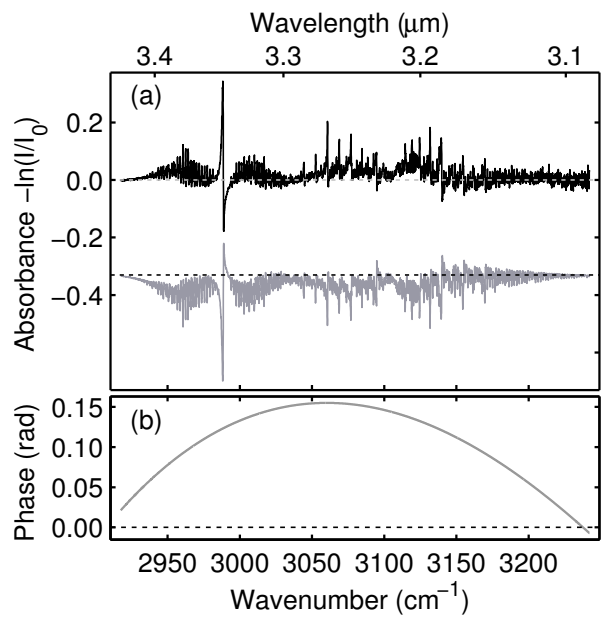}}
\caption{(a) Experimentally measured ethylene spectrum (black) at a concentration of 48 ppm and the corresponding calculated spectrum (gray). The calculated spectrum is inverted and offset for clarity. (b) Phase shift $\Delta\phi(\nu)$, which was used for the calculation in (a).}
\label{fig:Tobi_ethylene}
\end{figure}

A carefully controlled injection of 5.2 l of 1000 ppm ethylene in N$_2$ into the N$_2$-flushed OPO enclosure resulted in an estimated ethylene concentration of 48~ppm along the OPO path. Figure \ref{fig:Tobi_ethylene}(a) provides the obtained absorption spectrum, resulting from the injected ethylene taken within 60 s, with identical settings for the FTIR as described in Sec. \ref{sec:Ace_met}.
The simulated spectrum was obtained using the phase shift shown in Fig. \ref{fig:Tobi_ethylene}(b) and assuming a cavity round-trip loss of 25\%. We estimate a detection limit of 0.32 ppm for ethylene with our setup.
\section{Results with the Tm-pumped system}\label{sec:tm}
The Tm-system represents an attractive system because of its wide
bandwidth from 2.6--6.1 $\mu$m at the -30 dB level. Initial
measurements of intracavity spectra of water vapor and isotopic CO$_2$ in standard air have previously been presented in
Refs. \cite{mwh:cleo12, lein:Tm_spopo} for this system.
\subsection{Carbon monoxide (CO)}
Carbon monoxide absorbs around 4.7 $\mu$m, and the Tm-system was tuned
to this wavelength region by selecting an appropriate oscillation
peak, similar to the one shown in Fig. \ref{fig:Tm_spec}. The gas cell
was filled with 50 ppm CO in He and placed inside the OPO cavity. 
\begin{figure}[h!]
\centerline{\epsfbox{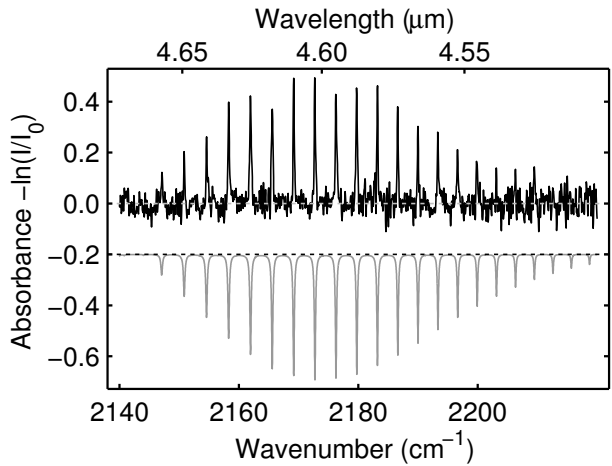}}
\caption{Measured (black) and calculated (gray) absorption spectra for 50~ppm carbon monoxide in helium at 1 atm. pressure. The calculated 
spectrum is offset and shown on an inverted scale for clarity. The effective path length is taken to be 7 times the length of the gas cell for 
the calculated spectrum.}
\label{fig:CO}
\end{figure}
Figure \ref{fig:CO} shows the measured and calculated absorption spectrum of CO, with a
measurement time of 2~min. In the calculations we assume a round-trip phase shift of $\Delta\phi=0$, which provided a
reasonable fit to the measured data. The detection limit for CO is estimated to be 0.27~ppm. The increased
noise properties of the Tm-system, compared to the Er-system, are
attributed to instabilities (mode hopping) in the 790-nm laser diodes used for pumping the Tm:fiber amplifier.
We observe that the intracavity absorption is enhanced by
a factor of $\sim 7$, corresponding to a round-trip loss of approximately $30\%$. However, this
is higher than the expected round-trip loss of approximately $20\%$, estimated in a previous publication of the Tm-system
\cite{lein:Tm_spopo}, and might be due to clipping of the signal beam
introduced by the intracavity gas cell. For a perfectly aligned gas cell, the clipping is
estimated to be a few percent.
\subsection{Isotopic carbon dioxide ($^{13}$CO$_2$)}
Isotopic ($^{13}$CO$_2$) carbon dioxide is naturally present in the atmosphere at concentration of approximately 4
ppm in volume \cite{wikipedia}, and can be easily detected using our Tm-laser based OPO without purging
the cavity. The resolution of the spectrometer in this experiment was 0.5 cm$^{-1}$ (we used model 80251 Oriel-Newport
mid-IR spectrometer). 
\begin{figure}[h!]
\centerline{\epsfbox{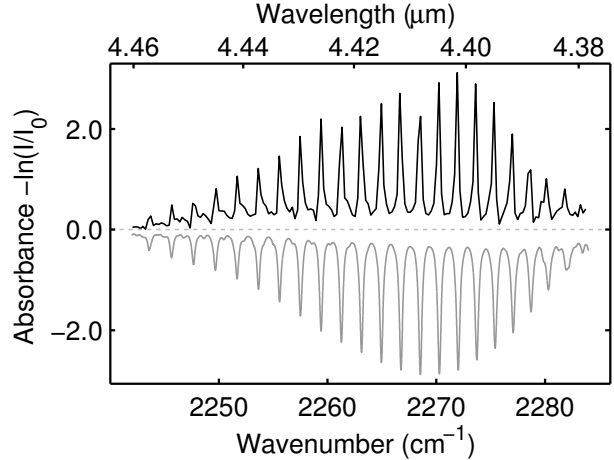}}
\caption{Measured (black) and calculated (gray) molecular spectra of
isotopic ($^{13}$CO$_2$) carbon dioxide. The simulation is based on the HITRAN database and is inverted for clarity.}
\label{fig:isotopic_CO2}
\end{figure}
Figure \ref{fig:isotopic_CO2} shows measured and calculated (HITRAN, with
matching spectral resolution of 0.5 cm$^{-1}$) spectra of isotopic carbon dioxide. To match the two
spectra in their peak values, we multiplied the HITRAN spectrum by 15 (assuming $\Delta\phi=0$), which means that our
cavity absorption enhancement factor is larger than the one we typically observe. This
discrepancy can be the result of an elevated (as compared to the average) amount of CO$_2$ in the
lab.
\section{Discussion}\label{sec:discussion}
The main motivation of using intracavity spectroscopy is to increase the
effective path length through the gas, without the need of
locking to an external Fabry-Perot cavity, or using an external
multi-pass cell. The effective path length in our experiments dependents on the round-trip
loss in the OPO cavity. We observe an
absorption enhancement up to a factor of 8, when the gas concentration was controlled, corresponding to a
round-trip loss of 25\%. This loss is comparatively high, and could be
due to clipping of the resonating beam in the OPO cavity. Based on the
mirror and coating specifications, a cavity round-trip loss of $5-10
\%$ should be obtainable, resulting in an enhancement factor
of $20-40$. Maximum effective path lengths are thus in the
range $100-200$ m. We emphasize that once the cavity round-trip
loss and dispersion has been determined (e.g. using a known gas with known concentration), both the enhancement factor and expected lineshapes 
are known for all molecules to be measured and thus absolute concentration measurements can be performed. 
\begin{table}[h]
\caption{Estimated detection limits.}
\begin{center}
\begin{tabular}{ccccc}\hline\label{tab:limits}
Gas name                               & Det. limit & Meas. time & Phys. path length    & System \\
                                       & [ppm]      & [s]        & [m]                  &        \\ \hline
Methane (CH$_4$)                       & 0.0017     & 57         & 3.75                 & Er     \\
Formaldehyde (CH$_2$O)                 & 0.31       & 30         & 0.48                 & Er     \\
Acetylene (C$_2$H$_2$)                 & 0.11       & 60         & 3.75                 & Er     \\
Ethylene (C$_2$H$_4$)                  & 0.32       & 60         & 3.75                 & Er     \\
Carbon monoxide (CO)                   & 0.27       & 120        & 0.48                 & Tm     \\
Isotopic carbon dioxide ($^{13}$CO$_2$) & 0.0024     & 90         & 4.00                 & Tm     \\ \hline
\end{tabular}
\end{center}
\end{table}
Table \ref{tab:limits} shows estimated detection limits for each of the six molecules investigated.
For calculating the molecular detection limit in our experiment, we have used the following procedure: First, we obtain the experimental set of 
points for the intracavity molecular absorbance $S(\nu_n)$. Second, from the known cavity dispersion and HITRAN database, we obtain the 
theoretical absorbance spectrum $T(\nu_n)$ -- for the same uniformly spaced set of points $\nu_n$. We remove the baseline from theoretical 
spectrum, since the baseline in 
our experiment is not well defined (we fit it to the valleys between absorption peaks). Next, we 'project' our experimental spectrum onto the 
theoretical one, that is we integrate the product  $S(\nu)T(\nu)$ over frequency, such that $I_1=\sum S(\nu_n)T(\nu_n)$. Then we estimate a 
standard deviation $\sigma$ (noise) of our experimental spectrum and computer-generate random 'noise'  $N(\nu_n)$ over the set $\nu_n$, with 
normal distribution and standard deviation $\sigma$ and project the noise onto the theoretical spectrum, i.e. integrate $N(\nu)T(\nu)$ over 
frequency and get $I_2=\sum N(\nu_n)T(\nu_n)$. The sum $I_2$ fluctuates around zero with each computer run, and by doing many ($>1000$) such 
runs, we get the standard deviation $\Delta I_2$. Finally, we calculate the minimal detectable concentration as the concentration used in the 
experiment, divided by a factor $I_1/\Delta I_2$ (which is an effective signal-to-noise factor). We think that sensitivity enhancement due to our 
'matched filter' approach is totally consisted with the 'multiple peaks' advantage described in Ref. \cite{adl:ftir_opo}.

During our experiments, the pump laser was free-running, and we
observed a drift of the OPO spectra on a time scale of the order of
minutes. This resulted in baseline drifts in the measured absorption
spectra. One improvement would be to lock the pump laser to an
external frequency reference. The OPO output would then be stabilized due to the self phase-locking mechanism
of degenerate OPOs \cite{wong:spopo}.
\section{Conclusions}\label{sec:conclusions}
By using synchronously pumped OPOs operating around degeneracy, we obtain ultra-broadband mid-IR radiation suitable for coherent spectroscopy 
in the Fourier domain. A large instantaneous bandwidth of up to 800 cm$^{-1}$ allows detection of several trace gases simultaneously. 
Spectroscopic detection of six molecules in trace amounts has been performed in the wavelength range of $2.5-5$ $\mu$m. 
By injecting the gases inside the OPO cavities we obtained substantial enhancement of the effective path length and achieved detection 
limits down to part-per-billion level in volume. Our intracavity sensing approach offers great simplicity and compactness, which might 
be a great asset for future applications. The dispersive spectral features that we observe at some circumstances are well reproduced 
using a simple model for propagation in a dispersive Fabry-Perot cavity. These features can be predicted a priori from knowledge of 
dispersion of the intracavity elements; on the other hand, if the dispersion of the individual components is not known, the overall dispersion 
can be precisely mapped by injecting trace amounts of known molecules and analyzing the shapes of spectral peaks. By decreasing the 
OPO loss and increasing the finesse of the OPO cavity we expect an improvement of detection limits, down to sub-ppb levels. This will also 
require better dispersion compensation, e.g. using chirped dielectric mirrors. Utilizing intracavity cells with low ($\sim$0.1 atm.) 
gas pressure will allow better specificity of molecular recognition because of sharper spectral features. With further development, this system 
may find important applications in trace gas detection and real time human breath analysis -- with further enhancement possible through the use 
of dual-comb multi-heterodyne methods.
\section*{Acknowledgments}
We greatly thank NASA, Office of Naval Research, Air Force Office of Scientific Research, Agilent Technologies, Sanofi-Aventis, 
Stanford University Bio-X, Stanford Medical School and Stanford Woods Institute for their financial support. We also thank Leo Hollberg, 
Robert Byer, Martin Fejer and Alex Kachanov for the most useful discussions. M.W.H. is indebted to Gunnar Arisholm for his comments on the 
manuscript.
\bibliographystyle{osajnl}

\begin{thebibliography}{10}
\newcommand{\enquote}[1]{``#1''}
\bibitem{esler:2000}
M.~B. Esler, D.~W.~T. Griffith, S.~R. Wilson, and L.~P. Steele,
  \enquote{Precision trace gas analysis by ft-ir spectroscopy. 1.
  {S}imultaneous analysis of {CO$_2$}, {CH$_4$}, {N$_2$O}, and {CO} in air,}
  Anal. Chem. \textbf{72}, 206--215 (2000).

\bibitem{Schliesser:comb}
A.~Schliesser, M.~Brehm, F.~Keilmann, and D.~W. van~der Weide,
  \enquote{Frequency-comb infrared spectrometer for rapid, remote chemical
  sensing,} Opt. Express \textbf{13}, 9029--9038 (2005).

\bibitem{thorpe:2008}
M.~J. Thorpe, D.~Balslev-Clausen, M.~S. Kirchner, and J.~Ye,
  \enquote{Cavity-enhanced optical frequency comb spectroscopy: application to
  human breath analysis,} Opt. Express \textbf{16}, 2387--2397 (2008).

\bibitem{ars:2011}
D.~D. Arslanov, K.~Swinkels, S.~M. Cristescu, and F.~J.~M. Harren,
  \enquote{Real-time, subsecond, multicomponent breath analysis by optical
  parametric oscillator based off-axis integrated cavity output spectroscopy,}
  Opt. Express \textbf{19}, 24078--24089 (2011).

\bibitem{risby:breath}
T.~H. Risby and F.~K. Tittel, \enquote{Current status of midinfrared quantum
  and interband cascade lasers for clinical breath analysis,} Opt. Eng.
  \textbf{49}, 111123 (2010).

\bibitem{didd:comb}
S.~A. Diddams, \enquote{The evolving optical frequency comb,} J. Opt. Soc. Am.
  B \textbf{27}, B51--B62 (2010).

\bibitem{didd:finger}
S.~A. Diddams, L.~Hollberg, and V.~Mbele, \enquote{Molecular fingerprinting
  with the resolved modes of a femtosecond laser frequency comb,} Nature
  \textbf{445}, 627--630 (2007).

\bibitem{sor:finger}
E.~Sorokin, I.~T. Sorokina, J.~Mandon, G.~Guelachvili, and N.~Picqu{\'{e}},
  \enquote{Sensitive multiplex spectroscopy in the molecular fingerprint 2.4
  $\mu$m region with a {C}r$^{2+}$:{ZnSe} femtosecond laser,} Opt. Express
  \textbf{15}, 16540--16545 (2007).

\bibitem{man:ftir}
J.~Mandon, G.~Guelachvili, and N.~Picqu{\'{e}}, \enquote{Fourier transform
  spectroscopy with a laser frequency comb,} Nature Photonics \textbf{3},
  99--102 (2009).

\bibitem{keil:dual}
F.~Keilmann, C.~Gohle, and R.~Holzwarth, \enquote{Time-domain mid-infrared
  frequency-comb spectrometer,} Opt. Lett. \textbf{29}, 1542--1544 (2004).

\bibitem{thorpe:spec}
M.~J. Thorpe and J.~Ye, \enquote{Cavity-enhanced direct frequency comb
  spectroscopy,} Appl. Phys. B \textbf{91}, 397--414 (2008).

\bibitem{bern:dual}
B.~Bernhardt, A.~Ozawa, P.~Jacquet, M.~Jacquey, Y.~Kobayashi, T.~Udem,
  R.~Holzwarth, G.~Guelachvili, T.~W. H{\"{a}}nsch, and N.~Picque,
  \enquote{Cavity-enhanced dual-comb spectroscopy,} Nature Phot. \textbf{4},
  55--57 (2009).

\bibitem{vaer:ftir}
X.~D.~D. Vaernewijck, K.~Didriche, C.~Lauzin, A.~Rizopoulos, M.~Herman, and
  S.~Kassi, \enquote{Cavity enhanced ftir spectroscopy using femto opo
  absorption source,} Mol. Phys. \textbf{109}, 2173--2179 (2011).

\bibitem{folty:spec}
A.~Foltynowicz, P.~Mas{\l}owski, A.~Fleisher, B.~Bjork, and J.~Ye,
  \enquote{Cavity-enhanced optical frequency comb spectroscopy in the
  mid-infrared application to trace detection of hydrogen peroxide,} Applied
  Physics B pp. 1--13 (2012).

\bibitem{till:meth}
K.~A. Tillman, R.~R.~J. Maier, D.~T. Reid, and E.~D. McNaghten,
  \enquote{Mid-infrared absorption spectroscopy of methane using a broadband
  femtosecond optical parametric oscillator based on aperiodically poled
  lithium niobate,} J. Opt. A: Pure Appl. Opt. \textbf{7}, S408--S414 (2005).

\bibitem{adl:opo}
F.~Adler, K.~C. Cossel, M.~J. Thorpe, I.~Hartl, M.~E. Fermann, and J.~Ye,
  \enquote{Phase-stabilized, 1.5 {W} frequency comb at 2.8--4.8 $\mu$m,} Opt.
  Lett. \textbf{34}, 1330--1332 (2009).

\bibitem{adl:ftir_opo}
F.~Adler, P.~Mas{\l}owski, A.~Foltynowicz, K.~C. Cossel, T.~C. Briles,
  I.~Hartl, and J.~Ye, \enquote{Mid-infrared {Fourier} transform spectroscopy
  with a broadband frequency comb,} Opt. Express \textbf{18}, 21861--21872
  (2010).

\bibitem{lein:spopo}
N.~Leindecker, A.~Marandi, R.~L. Byer, and K.~L. Vodopyanov, \enquote{Broadband
  degenerate opo for mid-infrared frequency comb generation,} Opt. Express
  \textbf{19}, 6296--6302 (2011).

\bibitem{lein:Tm_spopo}
N.~Leindecker, A.~Marandi, R.~L. Byer, K.~L. Vodopyanov, J.~Jiang, I.~Hartl,
  M.~Fermann, and P.~G. Schunemann, \enquote{Octave-spanning ultrafast opo with
  2.6-6.1 $\mu$m instantaneous bandwidth pumped by femtosecond {Tm}-fiber
  laser,} Opt. Express \textbf{20}, 7046--7053 (2012).

\bibitem{baev:intracavity}
V.~M. Baev, T.~Latz, and P.~E. Toschek, \enquote{Laser intracavity absorption
  spectroscopy,} Appl. Phys. B \textbf{69}, 171--202 (1999).

\bibitem{brunner:spectr}
W.~Brunner and H.~Paul, \enquote{The optical parametric oscillator as a means
  for intracavity absorption spectroscopy,} Opt. Comm. \textbf{19}, 253--256
  (1976).

\bibitem{boller:spectr}
K.-J. Boller and T.~Schr{\"{o}}der, \enquote{Demonstration of broadband
  intracavity spectroscopy in a pulsed optical parametric oscillator made of
  $\beta$-barium borate,} J. Opt. Soc. Am. B \textbf{10}, 1778--1784 (1993).

\bibitem{fol:2011}
A.~Foltynowicz, T.~Ban, P.~Mas{\l}owski, F.~Adler, and J.~Ye,
  \enquote{Quantum-noise-limited optical frequency comb spectroscopy,} Phys.
  Rev. Lett. \textbf{107}, 233002 (2011).

\bibitem{vod:spopo_gaas}
K.~L. Vodopyanov, E.~Sorokin, I.~T. Sorokina, and P.~G. Schunemann,
  \enquote{Mid-ir frequency comb source spanning 4.4--5.4 $\mu$m based on
  subharmonic {GaAs} optical parametric oscillator,} Opt. Lett \textbf{36},
  2275--2277 (2011).

\bibitem{kal:sol}
V.~L. Kalashnikov and E.~Sorokin, \enquote{Soliton absorption spectroscopy,}
  Phys. Rev. A \textbf{81}, 033840 (2010).

\bibitem{mar:spopo}
A.~Marandi, N.~C. Leindecker, V.~Pervak, R.~L. Byer, and K.~L. Vodopyanov,
  \enquote{Coherence properties of a broadband femtosecond mid-ir optical
  parametric oscillator operating at degeneracy,} Opt. Express \textbf{20},
  7255--7262 (2012).

\bibitem{dem:spec}
W.~Demtr{\"{o}}der, \emph{Laser Spectroscopy - Basic Concepts and
  Instrumentation} (Springer, 2003).

\bibitem{hitr:harvard}
\enquote{The {HITRAN} database,} http://www.cfa.harvard.edu/HITRAN/ .

\bibitem{gian:1999}
L.~Gianfrani, R.~W. Fox, and L.~Hollberg, \enquote{Cavity-enhanced absorption
  spectroscopy of molecular oxygen,} J. Opt. Soc. B \textbf{16}, 2347--2254
  (1999).

\bibitem{mwh:cleo12}
M.~W. Haakestad, N.~Leindecker, A.~Marandi, J.~Jiang, I.~Hartl, M.~Fermann, and
  K.~L. Vodopyanov, \enquote{Broadband intracavity molecular spectroscopy with
  a degenerate mid-{IR} {OPO},} in \enquote{{C}onference on {L}asers and
  {E}lectro-{O}ptics ({CLEO}),}  (San Jose, California, USA, 2012). Paper no.
  CF2C.2.

\bibitem{wikipedia}
\enquote{Wikipedia,} http://en.wikipedia.org/ .

\bibitem{wong:spopo}
S.~T. Wong, K.~L. Vodopyanov, and R.~L. Byer, \enquote{Self-phase-locked
  divide-by-2 optical parametric oscillator as a broadband frequency comb
  source,} J. Opt. Soc. Am. B \textbf{27}, 876--882 (2010).

\end{thebibliography}

\end{document}